\documentclass[12pt]{article}
\usepackage[utf8]{inputenc}
\usepackage{geometry}
\usepackage{graphicx}
\usepackage{float}
\usepackage{indentfirst}
\usepackage{amsfonts, amsmath, amssymb, amsthm}
\usepackage[labelfont=bf]{caption}
\usepackage{booktabs}
\usepackage{chemfig}
\usepackage{lipsum}
\usepackage{verbatim}
\usepackage{csquotes}
\usepackage[natbib=true, style=abnt, justify]{biblatex}
\usepackage[T1]{fontenc}
\usepackage{calligra}

\title{Application of Molecular Topology to the Prediction of Antioxidant Activity in a Group of Phenolic Compounds}

\author{Jaime Barros Silva Filho\\ 
Fernando de Souza Bastos\\ 
Diogo da Silva Machado\\
Maria Luiza Ferreira Delfim}

\begin{document}
\maketitle
\begin{abstract}
The study of compounds with antioxidant capabilities is of great interest to the scientific community, as it has implications in several areas, from Agricultural Sciences to Biological Sciences, including Food Engineering, Medicine and Pharmacy. In applications related to human health, it is known that antioxidant activity can delay or inhibit oxidative damage to cells, reducing damage caused by free radicals, helping in the treatment, or even preventing or postponing the onset of various diseases. Among the compounds that have antioxidant properties, there are several classes of Phenolic Compounds, which include several compounds with different chemical structures. In this work, based on the molecular branching of compounds and their intramolecular charge distributions, and using Molecular Topology, we propose a significant topological-mathematical model to evaluate the potential of candidate compounds to have an antioxidant function.
\end{abstract}

\section{Introduction}

Studies and research on compounds with antioxidant activities have gained prominence in recent years. This is due to the antioxidant activity delay or inhibiting the oxidative damage of cells, decreasing the damage caused by free radicals, preventing or postponing the onset of various diseases and also helping in the treatment of diseases such as diabetes, inflammatory diseases, cardiovascular disorders, associated with oxidative stress, some types of cancer and aging (see eg \citet{XX1, ex1, ex4, ex5, ex2, ex3}).  

Measuring the antioxidant activity/capacity of foods and biological compounds is therefore essential not only to ensure food quality, but also to assess the efficiency of dietary antioxidants in the prevention and treatment of diseases related to oxidative stress.

Among the compounds that have antioxidant properties, there are several classes of Phenolic Compounds, which are those that have one or more hydroxyl groups directly linked to an aromatic ring. Such compounds represent a large group of molecules with a variety of functions related to plant growth, development and defense. They include signaling molecules, pigments and flavors that can attract or repel, as well as compounds that can protect the plant against insects, fungi, bacteria and viruses.

With the growing recognition of their bioactivity values, the research effort on the antioxidant behavior of Phenolic Compounds, which include thousands of compounds with different chemical structures, has increased significantly in recent decades. Furthermore, the methods and instruments used to measure antioxidant activity/capacity have made remarkable progress. According to \citet{activity}, there are several methods for assessing this ability, which generally fall into three distinct categories, namely, spectrometry, electrochemical assays and chromatography. However, all are based on chemical reactions that demand time and financial resources. 

Therefore, the implementation of mechanisms that can provide an early estimate of the potential of a molecule to exert an antioxidant function is an interesting research strategy. Molecular Topology based on Quantitative Structure Activity Relations (QSAR) is a tool that has great potential in this context. The QSAR technique, developed by \citet{hansch1964p}, combines Topology and Statistics concepts and its approach can be described as a statistical method of data analysis to develop models that can correctly predict certain biological activity or properties of compounds based on their structure chemistry. QSAR techniques apply descriptors based on molecular structures and use algorithms to correlate the obtained descriptors with the value of the target property of interest, for more information we suggest \citet{cramer2012inevitable}.

The development of Molecular Topology is attributed to Randic \cite{XX2} and Kier-Hall \cite{XX4, XX3} and is based, above all, on the premise that, in many cases, there is a close relationship between the structures of organic compounds and many of its chemical and biological properties. From which, a set of suitable numerical characterizations for the molecules of interest is obtained. Such characterizations are designated by topological indices, which allow, after statistical treatment (specifically, Linear Discriminant Analysis), the classification into groups, with greater and lesser probabilities of having a chemical function of interest. It should be noted that Molecular Topology has been used satisfactorily in obtaining new compounds for the production of drugs, cosmetics and agrochemicals of great interest to society, see more details in \citet{XX6, XX7} and \citet{XX5}.

To obtain the topological indices, the molecules to be considered are mathematically modeled as Topological Graphs, naturally taking their atoms as vertices and atomic bonds as edges. Thus, from the graphs, numerical matrices are obtained containing information about the molecules (such as the structural arrangement, global charge transfer and between pairs of atoms, atomic electronegativity, among others), which, through specific formulas, lead to the survey of topological indices.

The main objective of this study was to develop a predictive QSAR model to investigate the antioxidant activity of a group of Phenolic Compounds using Molecular Topology and Linear Discriminant Analysis.

\section{Materials and methods}

\subsection{Compounds Analyzed and Their Antioxidant Activities}

In this work we consider a set of 23 Phenolic Compounds, whose structures can be seen in Tables \ref{tab1}, \ref{tab2} and \ref{tab3}. 

\begin{table}[H]
\centering
\begin{tabular}{|c|c|c|c|}
\hline
&&&\\
\chemfig[atom style={scale=0.3, rotate=80}]{[:30]*6((-H3C)-(-CH3)=(*6(-O-(-[:-30]CH3)(-[:30](-[:-30](-[:30](-[:-30](-[:-90]CH3)(-[:30](-[:-30](-[:30](-[:-30](-[:-90]CH3)(-[:30](-[:-30](-[:30](-[:-30](-[:30]CH3)(-[:-90]CH3)))))))))))))---))-=(-H3C)-(-HO)=)}&
\chemfig[atom style={scale=0.3, rotate=80}]{[:30]*6(-(-CH3)=(*6(-O-(-[:-30]CH3)(-[:30](-[:-30](-[:30](-[:-30](-[:-90]CH3)(-[:30](-[:-30](-[:30](-[:-30](-[:-90]CH3)(-[:30](-[:-30](-[:30](-[:-30](-[:30]CH3)(-[:-90]CH3)))))))))))))---))-=(-H3C)-(-HO)=)} &
\chemfig[atom style={scale=0.3, rotate=80}]{[:30]*6((-H3C)-(-CH3)=(*6(-O-(-[:-30]CH3)(-[:30](-[:-30](-[:30](-[:-30](-[:-90]CH3)(-[:30](-[:-30](-[:30](-[:-30](-[:-90]CH3)(-[:30](-[:-30](-[:30](-[:-30](-[:30]CH3)(-[:-90]CH3)))))))))))))---))-=-(-HO)=)}&
\chemfig[atom style={scale=0.3, rotate=80}]{[:30]*6(-(-CH3)=(*6(-O-(-[:-30]CH3)(-[:30](-[:-30](-[:30](-[:-30](-[:-90]CH3)(-[:30](-[:-30](-[:30](-[:-30](-[:-90]CH3)(-[:30](-[:-30](-[:30](-[:-30](-[:30]CH3)(-[:-90]CH3)))))))))))))---))-=-(-HO)=)}\\
&&&\\
\scriptsize{1 - $\alpha$-Tocoferol}&
\scriptsize{2 - $\beta$-Tocoferol} &
\scriptsize{3 - $\gamma$-Tocoferol}&
\scriptsize{4 - $\delta$-Tocoferol}\\
\hline
&&&\\
\chemfig[atom style={scale=0.5}]{*6((-O(-[:150]H3C))-(-OH)=(-O(-[:30]CH3))-=(-(-[:30]OH)(=[:150]O))-=-)}&
\chemfig[atom style={scale=0.4}]{*6((-HO)-=(-OH)-=(-(=[:30](-[:90]*6(-=-(-OH)=-=))))-=-)}&
\chemfig[atom style={scale=0.4}]{*6((-HO)-(-OH)=(-OH)-=(-(-[:30]OH)(=[:150]O))-=-)}&
\chemfig[atom style={scale=0.4}]{*6((-HO)-=-(-OH)=(-(-[:30]OH)(=[:150]O))-=-)}\\
&&&\\
\scriptsize{5 - Siringico}& 
\scriptsize{6 - Reverastrol}& 
\scriptsize{7 - Ácido Galico}& 
\scriptsize{8 - Ácido Gentissico}\\
\hline
\multicolumn{4}{|c|}{}\\
\multicolumn{4}{|c|}{
\chemfig[atom style={scale=0.4, rotate=160}]{*6((-HO)=(-O(-[:-30]CH3))-=(-(=[:90](-(-[:-30](-[:30](=[:90]O)(-[:-30](=[:30](-[:-30](*6(=-=(-OH)-(-O(-[:-30]CH3))=-)))))))(=[:90]O))))-=-)}
}\\
\multicolumn{4}{|c|}{\scriptsize{5 - Curcumina}}\\
\hline
\end{tabular}
\caption{Structural formulas of Phenolic Compounds used in the present study with antioxidant capacity.}
\label{tab1}
\end{table}

\begin{table}[H]
\centering
\begin{tabular}{|c|c|c|c|}
\hline
&&&\\
\chemfig[atom style={scale=0.5}]{*6(-(-OH)=-=(-(=[:150]-[:90](-[:30]OH)(=[:150]O)))-=-)}&
\chemfig[atom style={scale=0.5}]{*6(-(-OH)=(-O(-[:30]CH3))-=(-(-[:30]OH)(=[:150]O))-=-)}&
\chemfig[atom style={scale=0.5}]{*6(-(-OH)=(-O(-[:30]CH3))-=(-(=[:150]-[:90](-[:30]OH)(=[:150]O)))-=-)}&
\chemfig[atom style={scale=0.5}]{*6(-(-OH)=(-O(-[:30]CH3))-=(-(=[:30]O))-=-)}\\
&&&\\
\scriptsize{10 - Ácido $p$-cumarico}&
\scriptsize{11 - Ácido Vanilico} & 
\scriptsize{12 - Ácido Ferulico}&
\scriptsize{13 - Vanilina}\\
\hline
&&&\\
\chemfig[atom style={scale=0.5}]{*6(-=-(-OH)=(-(=[:150]-[:90](-[:30]OH)(=[:150]O)))-=-)}&
\chemfig[atom style={scale=0.5}]{*6(-(-OH)=-=(-(-[:30]OH)(=[:150]O))-=-)}&
\chemfig[atom style={scale=0.5}]{*6(-(-OH)=(-OH)-=(-(-[:150](-[:90]OH)(=[:210]O)))-=-)}&
\chemfig[atom style={scale=0.5}]{*6(-=-(-OH)=(-(-[:30]OH)(=[:150]O))-=-)}\\
&&&\\
\scriptsize{14 - Ácido $o$-cumarico}&
\scriptsize{15 - Ácido $p$-hidroxibenzoico}&
\scriptsize{16 - Ácido Protocatequino}&
\scriptsize{17 - Ácido Salicilico}\\
\hline
\multicolumn{4}{|c|}{}\\
\multicolumn{4}{|c|}{
\chemfig[atom style={scale=0.5, rotate=80}]{*6(-(-OH)=(-O(-[:30]CH3))-=(-(-[:30]-[:90](-[:30]CH3)(=[:150]O)))-=-)}
}\\
\multicolumn{4}{|c|}{\scriptsize{18 - Gengerona Zigerona}}\\
\hline
\end{tabular}
\caption{Structural formulas of Phenolic Compounds used in the present study, classified with non-active antioxidant capacity.}
\label{tab2}
\end{table}

We used topological indices, described in the subsection \ref{indices}, related to the compounds in Tables \ref{tab1} and \ref{tab2} as vectors of observed characterists to estimate the discriminant function. There are several indices in the literature, each one more sensitive to a certain characteristic of the compounds. Table \ref{tab1} contains all compounds that have proven antioxidant capacity. Table \ref{tab2} has the same amount of phenolic compounds as Table \ref{tab1}, however, such compounds either have low or no antioxidant activity.

In Table \ref{tab3} we present a group of five phenolic compounds with proven antioxidant capacity, such compounds were used to apply the discriminant function, that is, they are our test group.

\begin{table}[H]
\centering
\begin{tabular}{|c|c|c|c|}
\hline
&&&\\
\chemfig[atom style={scale=0.5, rotate=80}]{[:30]*6((-HO)-=(*6(-O-(-(*6(=-=(-OH)-(-OH)=-)))-(-OH)--))-=(-HO)-=)}&
\chemfig[atom style={scale=0.5, rotate=80}]{*6((-HO)=(-OH)-=(-(=[:-30](-[:30](=[:90]O)(-[:-30]O(-[:30]*6(-(-OH)-(-OH)-(-[:60]OH)(-[:0](=[:60]O)(-[:-60]OH))---))))))-=-=)}&
\chemfig[atom style={scale=0.5, rotate=80}]{*6((-HO)-=-(-(-[:-30](-O(-[:-30](-(=[:-30](-(*6(=-=(-OH)-(-OH)=-)))))(=[:-90]O)))(-[:-90](-[:-30]OH)(=[:-150]O))))=-(-HO)=)}&
\chemfig[atom style={scale=0.5, rotate=80}]{*6((-HO)=(-OH)-(*6(-O-(=O)-(*6(=-(-OH)=(-OH)-=-))---))=(*6(---O-(=O)-))-=-=)}\\
&&&\\
\scriptsize{19 - Catequina}&
\scriptsize{20 - Ácido Clorogenico}&
\scriptsize{21 - Ácido Rosmanico}& 
\scriptsize{22 - Ácido Elagico}\\
\hline
\multicolumn{4}{|c|}{}\\
\multicolumn{4}{|c|}{
\chemfig[atom style={scale=0.5, rotate=-20}]{*6(-(-OH)=(*6(-(=O)-(-OH)-(-(*6(=-=(-OH)-(-OH)=-)))-O-))-=-(-HO)=)}
}\\
\multicolumn{4}{|c|}{\scriptsize{23 - Quercetina}}\\
\hline
\end{tabular}
\caption{Structural formulas of Phenolic Compounds used in the present study as a test group, all with antioxidant capacity.}
\label{tab3}
\end{table}


\subsection{Calculation of Topological Indices} \label{indices}

Topological indices are numerical functions of molecular graphs and are considered important molecular descriptors. According to \citet{vasilyev2014mathchem}, in addition to being graph invariant, topological indices do not consider information about molecular geometry, such as bond lengths, bond angles or twist angles, but instead encode information about adjacencies of atoms and branches within a molecule. Also according to the same authors, since the computation of topological indices uses less resources than the computation of those molecular descriptors that also take into account molecular geometry, topological indices have gained considerable popularity and many new topological indices have been proposed and studied in the literature specialized in recent years.

We assume that the following aspects of molecules are relevant to the investigation of the occurrence of antioxidant activity: the type of molecular branching and the distribution of intramolecular charge. Therefore, in this work, we use two types of topological indices: the {\it Randic index} $\chi^1$ and the {\it topological load indices} $G_4, J_2$ and $J_5$.

All the topological indices used in this work were obtained with the aid of an appropriate python algorithm that is available in the supplementary material that accompanies this article.

\subsubsection{Randic Index}

According to \citet{GUTMAN2018307}, among the several hundred descriptors of molecular structures based on graphs (see \citet{todeschini2008handbook}), the Randic index is certainly the most widely applied in chemistry and pharmacology. This index characterizes the branching of the molecular graph and was introduced by Milan Randic \cite{XX2}. If $E(G)$ denotes the set of edges of the molecular graph $G$, then the Randic index of $G$ is defined by
\begin{align}
\chi^1(G) = \displaystyle\sum_{e_{ij}\in E(G)} (deg_i\cdot deg_j)^{-1/2},
\end{align}
where, $deg_i$ and $deg_j$ are the degrees of vertices $i$ and $j$, respectively.

\subsubsection{Topological Load Indexes}

Topological Charge Indices were introduced in the literature by \citet{XX66} and have the ability to describe molecular charge distribution. In fact, given a molecular graph $G$, let $A(e)$ be its adjacency matrix (modified), with the relative electronegativity of each atom in the main diagonal entries and $D^{\ast}$ the distance matrix inverse square of $G$, with entries on the main diagonal taking on a value of zero. Consider $M = [m_{ij}]$ the square matrix of order $N$ (where $N$ is the number of vertices of $G$) defined by
\begin{align}
    M = A(e) \times D^{\ast}
\end{align}
and take, for each $i,j$, with $1\leq i,j \leq N$ the {\it charge term} $CT_{ij}$ defined by
\begin{align}
    CT_{ij} = m_{ij} - m_{ji}.
\end{align}
The (valency) {\it topological load indices}, $G_k$, with $1 \leq k \leq N-1$ is defined by
\begin{align}
    G_k = \sum^{i = N-1, j = N}_{i,j=i+1}\delta(k, d_{ij}) \, \cdot \mid CT_{ij}\mid,
\end{align}
where $\delta$ is the Kronecker delta function ($\delta(p,q) = 1$, if $p = q$ and $0$ otherwise) and $d_{ij}$ denote the entries of the matrix of topological distance. Remember that in the main diagonal entries of the matrix $A(e)$ the relative electronegativity of an element $Q$ can be calculated by the formula
\begin{align}
    Re^-(Q) = \lambda\cdot(e^-(Q) - e^-(C)), 
\end{align}
where $e^-(Q)$ and $e^-(C)$ denote, respectively, the Pauling electronegativity of the element $Q$ and of the carbon atom $C$. The value $\lambda$ constitutes the conversion factor. In this work, we consider two values for $\lambda$: 2.2 (using hydrogen normalization = 0.77) and 3.28 (using chlorine normalization = 2). For this reason, to distinguish the two respective situations, we will use the symbols $G_k^{2.2}$ and $G_k^{3.28}$ to indicate the considered conversion factor.

On the other hand, the index $J_k$ is defined by,
\begin{align}
    J_k = \frac{G_k}{N-1},\quad k=1,\cdots, N-1
\end{align}
which measures the average value of charge transfer for each chemical bond in the substance.

\bigskip

\begin{table}[H]
\centering
\begin{tabular}{lrrrrrrrl}
\hline
\toprule
\emph{Comp. no.} & \emph{$\chi^1$} &\emph{$G_4^{3.28}$}&\emph{$J_2^{3.28}$}& \emph{$J_5^{3.28}$} &  \emph{$G_4^{2.2}$} & \emph{$J_2^{2.2}$} & \emph{$J_5^{2.2}$}\\
\hline
\toprule
& \multicolumn{6}{c}{Antioxidant group training}\\
\hline
\toprule
{\bf 1} &	13,72456 &	3,59624 &	0,33965 &	0,05315 &	3,41601	& 0,33924 & 0,05059\\  	
{\bf 2} &	13,19576 &	3,04046 &	0,3174 &	0,05595 &	2,92031 &	0,31262	& 0,05329 \\
{\bf 3} &	13,19576 &	3,07157	& 0,3174 &	0,05666	& 2,95142 &	0,31262 &	0,05401 \\ 
{\bf 4} &	12,77311 &	2,51578	& 0,26956 &	0,05174	& 2,45571 &	0,2603 &	0,049\\ 
{\bf 5}	& 5,35317 &	1,55516	& 0,41686 &	0,06137 &	1,14092	& 0,4076 &	0,04784\\
{\bf 6} &	5,63077 &	0,59578 &	0,26229 &	0,01284 &	0,53571 &	0,3079 &	0,00993\\ 
{\bf 7} &	4,37239 &	0,67669	& 0,46465 &	0,02008	& 0,59762 &	0,51897 &	0,01968\\ 
{\bf 8} &	3,84159 &	0,84513	& 0,42525 &	0,0125 &	0,73126 &	0,43543 &	0,0125\\
{\bf 9} &	9,83297	& 2,81071 &	0,5237 &	0,06254 &	2,5695 &	0,51444 &	0,04775\\
\hline
\toprule
& \multicolumn{6}{c}{Non-antioxidant group training}\\
\hline
\toprule
{\bf 10} &	4,0856 &	0,57801	& 0,54821 &	0,04045 &	0,51793 &	0,52525 &	0,02297\\
{\bf 11} &	4,40225 &	1,18803 &	0,4169 &	0,03334 &	1,01364 &	0,40595 &	0,03293\\ 
{\bf 12}	& 5,06293 &	1,14313 &	0,50661 &	0,05282	& 1,102253 &	0,49734 &	0,03508 \\ 
{\bf 13} &	3,99562 &	1,17424	& 0,34116 &	0,0303 &	1,05364 &	0,35315 &	0,025\\
{\bf 14} &	3,89175 &	0,76889 &	0,42424	& 0,03879 &	0,76889 &	0,45645 &	0,03237 \\ 
{\bf 15} &	3,42492 & 	0,71179 &	0,44781	& 0,01066 &	0,53157 &	0,41975 &	0,01016\\ 
{\bf 16} &	4,29869 &	1,01402 &	0,45592 &	0,0397 &	0,83379 &	0,47666 &	0,02371 \\ 
{\bf 17} &	3,4309 &	0,74291 &	0,45847 &	0,00694 &	0,56268 &	0,44444 &	0,00694\\ 
{\bf 18} & 5,34935	& 1,43179 &	0,47242 &	0,06691	& 1,37126 &	0,46315 &	0,05508 \\
\hline
\toprule
& \multicolumn{6}{c}{Test group}\\
\hline
\toprule
{\bf 19} &	8,1157 &	1,84068	& 0,31414 &	0,04914 &	1,84651 &	0,33093 &	0,03719 \\ 
{\bf 20} &	9,89595 &	2,43829 &	0,52694 &	0,03863 &	1,83233 &	0,43631 &	0,02781\\ 
{\bf 21} &	9,40614 &	2,59738 &	0,50404	& 0,06857 &	2,11678 &	0,50306 &	0,04338\\ 
{\bf 22} &	8,02602 &	3,50632	& 0,47763 &	0,05734 &	2,66436	& 0,5331 &	0,04339\\ 
{\bf 23} &	7,76622 &	2,52803 &	0.33478	& 0,04599 &	246.000	& 0.40183 &	0,04106\\

\bottomrule 
\hline
\end{tabular}
\caption{Topological indices related to Phenolic Compounds presented in Tables \ref{tab1} and \ref{tab2} broken down by group, as used in the Discriminant Analysis.}
\label{tab4}
\end{table}

\bigskip

\subsection{QSAR Algorithms: Linear Discriminant Analysis}

With the objective of detecting the presence of the antioxidant function in Phenolic Compounds, the QSAR prediction model by Linear Discriminant Analysis (LDA) was established, a branch of multivariate statistics used in problems of discrimination and classification of categories or objects, and whose idea base appear in R. A. Fisher's seminal work \cite{XX8}, and are currently a widespread topic (see \citet{XX9, XX10}). In addition, several works use Discriminant Analysis in establishing QSAR models, see, for example, \citet{qsar1, qsar2, qsar3, qsar4, qsar5}.

Although Discriminant Analysis cannot provide concrete predicted values of the antioxidant effect, it can determine the likelihood of classifying compounds as active or inactive for antioxidant activity and thus aid in the discovery and development of efficient antioxidants. Therefore, we obtained a discriminant function (DF) that allows a classification between compounds with active and inactive antioxidant activity. This function linearly depends on the topological indices considered, which are assumed to be usable in distinguishing compounds.

In the set of Phenolic Compounds used to obtain the discriminant function, we separated each compound into two groups: the first formed by compounds with proven antioxidant activity and the second by compounds without antioxidant activity. The LDA was applied using the statistical software \citet{softwareR}.

\section{Results and Discussion}

The Discriminant Analysis was applied to a training group formed by 18 compounds, of which 09 are classified as active for the antioxidant function and 09 are classified as non-active for this function. For classification, experimental data were taken from the literature and consisted of antioxidant activity determined by DPPH free radical scavenging using lipid peroxidation inhibitory effects expressed as the concentration of $50\%$ lipid peroxidation inhibition ($IC_{50}/\ mu mol\ , L^-1$). In addition, we considered a test group formed by 05 Phenolic Compounds, with active antioxidant function, presented in Table \ref{tab4}. All phenolic compounds considered in this work are found in foods used in the human diet.

As expected, the discriminant function (DF) found depends linearly on its variables, which are given by each of the topological indices used in the study:

\begin{align*}
{\bf DF} &= -31.023674 + 4.4419931\, {\bf \chi^1} +41.2828013\, {\bf G_4^{3.28}} - 151.5908411\, {\bf J_2^{3.28} } - \\
&- 270.9518622\, {\bf J_5^{3.28}} - 52.570601\, {\bf G_4^{2.2}} + 180.1409685\, {\bf J_2^{2.2}} + 310.4418446\, {\bf J_5^{2,2}}
\end{align*}

The \textbf{DF} is used to characterize the presence of antioxidant function in compounds: a certain compound will be selected as potential antioxidant if $DF\geq 31.02367$, otherwise, if $DF<31.02367$, then it is classified as not antioxidant.

The most important fact about the $DF$ function is determining its precision. For this, we apply the discriminant function on the outer test group. The result was: $80\%$ precision. Table \ref{tab5} presents the results obtained for each compound. In the test group compounds, only 01 compound (Catechin) was incorrectly classified by the $DF$ function, as non-oxidant ($DF<31.02367$). For the training set, sorting by the $DF$ function was also very significant. In fact, $88.9\%$ correct prediction for the antioxidant group, 8 out of 9 classified correctly. In the non-antioxidant group, $100\%$ of correct prediction, according to Table \ref{tab5}.

\begin{table}[H]
\centering
\begin{tabular}{cccc}
  \hline
    & Compostos        & DF & Classificacao \\ 
  \hline
  1 & Catequina        & 25.19& Non\-Antioxidant \\ 
  2 & Ácido Clorogenico& 45.18 & Antioxidant \\ 
  3 & Ácido Rosmanico  & 46.83 & Antioxidant \\ 
  4 & Ácido Elagico    & 61.90 & Antioxidant \\ 
  5 & Quercitina       & 31.46 & Antioxidant \\ 
  6 & Alphatacoferol   & 40.77 & Antioxidant \\ 
  7 & Betatacoferol    & 40.20 & Antioxidant \\ 
  8 & Gammatacoferol   & 39.88 & Antioxidant \\ 
  9 & Deltatacoferol   & 38.72 & Antioxidant \\ 
  10 & Curcumina       & 35.79 & Antioxidant \\ 
  11 & Reverastrol     & 36.75 & Antioxidant \\ 
  12 & Ácido Galico    & 39.66 & Antioxidant \\ 
  13 & Ácido Gentisico & 27.98 & Non\-Antioxidant \\ 
  14 & Ácido Sirigico  & 36.46 & Antioxidant \\ 
  15 & Ácido Pcumarico & 22.47 & Non\-Antioxidant \\ 
  16 & Ácido Vanilico  & 26.43 & Non\-Antioxidant \\ 
  17 & Ácido Ferulico  & 21.11 & Non\-Antioxidant \\ 
  18 & Vanilina & 22.29& Non\-Antioxidant \\ 
  19 & Ácido Ocumarico & 26.06 & Non\-Antioxidant \\ 
  20 & Ácido Phidroxibenzoico & 24.65 & Non\-Antioxidant \\ 
  21 & Ácido Protocatequino & 30.48 & Non\-Antioxidant \\ 
  22 & Ácido Salicilico & 27.17 & Non\-Antioxidant \\ 
  23 & Gengerona Zigerona & 21.57 & Non\-Antioxidant \\ 
   \hline
\end{tabular}
\caption{Classification of compounds after application of the Discriminant Function.}
\label{tab5}
\end{table}

\section{Conclusions}

Discriminant Analysis was applied to establish the QSAR model incorporating Topological Indices that proved satisfactory to predict the antioxidant activity of 23 known Phenolic compounds. We made use of topological indices, which are particularly sensitive to two types of molecular information, namely, the molecular branching of the compound that is captured by $\chi^1$ and the intramolecular charge distributions of the compound through charge indices $G_4 , J_2 , J_5$. We believe that our methodology can be applied to assess the probability of existence of antioxidant activity in new molecular compounds, before the application of chemical methods, saving researchers time and financial resources.

\section{acknowledgment}
\noindent The third author thanks the University of California Riverside for the hospitality.

\printbibliography

\newpage
\section{Supplementary Material} \label{AAA}

\begin{verbatim}
\noindent import networkx as nx\\
\noindent import numpy as np\\
\noindent import matplotlib.pyplot as plt\\
\noindent import networkx.drawing\\
\noindent import math
\medskip
\noindent def mod(x):
  if x>0 or x==0:
     return x
  elif x <0:
     return -x

def find_shortest_path(graph,start,end, path=[]):
  path=path + [start]
  if start == end:
    return path
  shortest= None
  for node in graph[start]:
    if node not in path:
      newpath=find_shortest_path(graph, node, end, path)
      if newpath:
          if not shortest or len(newpath) < len(shortest):
              shortest=newpath
  return shortest

def Randic(MA,GRAU):
  r=0
  for i in range(0,n):
    for j in range(i,n):
      if MA[i][j]!=0 and GRAU[i]!=0 and GRAU[j]!=0:
        r = r + 1/(GRAU[i]*GRAU[j])**0.5
      else:
        r=r
  return r

def galvez(k,MD,MC, c=0):
    for i in range(0,n):
      for j in range(i,n):
        if k == MD[i][j]:
           c = c + mod(MC[i][j])
    return c


def Matriz_Adjacencia(n):
    MAD=np.zeros((n,n))
    i=0
    j=0
    resp=0
    print()
    print('ATENÇÃO APENAS OS NUMEROS ACIMA DA DIAGONAL PRINCIPAL (MAij TEM J>I) 
    E A ELETRONEGATIVIDADE NA DIAGONAL PRINCIPAL')
    print()
    while True: 
        i= int(input('qual o índice da linha?')) - 1
        while i>n-1 or i=='':
              print('digite um valor menor que %d' %(n+1))
              i=int(input('qual o índice da linha?')) - 1
        j=int(input('qual o índice da coluna?')) - 1
        while j>n-1 or j=='':
            print('digite um valor menor que %d' %(n+1))
            j=int(input('qual o índice da coluna?')) - 1
        MAD[i][j]=  float(input('qual o valor?'))
        if MAD[i][j] == 0:
           resp = input('voce deseja colocar mas algum valor?')
           if resp == 'não' or resp == 'nao' or resp=='n':
               break

    

    for i in range(0,len(MAD)):
      for j in range (i,len(MAD)):
        MAD[j][i]=MAD[i][j]
    return MAD

def print_matriz(M):
    print()
    for i in range(0,len(M)):
        for j in range(0,len(M)):
            print('%2.4f' %M[i][j], end=' ')
        print()

def verificar(M):
    pup=0
    for i in range(0,len(M)):
      for j in range(0,len(M)):
        if M[i][j]==0:
            pup=pup+1
      if pup == len(M):
          print('os valores da linha/coluna ', i+1,' estao errados. Digite novamente')
          MAD=Matriz_Adjacencia(len(M))
      pup=0

#MONTAR MATRIZ ADJACÊNCIA MODIFICADA:

n=int(input('numero de elementos do grafo'))
MAM= Matriz_Adjacencia(n)

for i in range(0,len(MAM)):
    for j in range (0,len(MAM)):
        if i == j and MAM[i][j]!=0:
            MAM[i][j]= 3.28*(MAM[i][j]-2.55)
        else:
            MAM[i][j]=MAM[i][j]
print()
print('MATRIZ DE ADJACÊNCIA MODIFICADA 3,28:')
print()
print_matriz(MAM)

#MONTAR MATRIZ DE ADJACÊNCIA:

MA=np.zeros((n,n))
for i in range(0,n):
    for j in range(0,n):
        if i!=j:
            MA[i][j]=MAM[i][j]
        else:
            MA[i][j]=0

#MONTAR DICIONARIO DE LISTA DE ADJACENCIA:

M=np.zeros((n,n))
for i in range(0,n):
 p=0
 for j in range(0,n):
  if MA[i][j]!=0:
      M[i][p]=j
      p=p+1
          
print()

dicio={}
lista=[]
for i in range(0,n):
    for j in range(0,n):
       if M[i][j]!=0 or j==0:
          lista.append(M[i][j])
       if j==n-1:
         dicio[i]=lista[:]
         lista=[]

#MONTAR MATRIZ DE DISTÂNCIA

MD=np.empty((n,n))

for i in range(0,n):
  for j in range(0,n):
        MD[i][j]=len(find_shortest_path(dicio,i,j)) -1

print()
print('MATRIZ DISTÂNCIA:')
print()
print_matriz(MD)

#MONTAR MATRIZ DISTÂNCIA MODIFICADA:

MDM=np.zeros((n,n))

for i in range(0,n):
  for j in range(0,n):
      if MD[i][j]!=0:
         MDM[i][j]= 1/((MD[i][j]))**2

#CALCULAR O GRAU DOS VERTICES:

GRAU=np.zeros(n)

for i in range (0,n):
    for j in range(0,n):
      GRAU[i]= GRAU[i]+MA[i][j]

#MUlTIPLICAR A MATRIZ CRIANDO MATRIZ INTERMEDIARIA: 

Mint=np.empty((n,n))

for i in range(0,n):
    p=0
    for j in range(0,n):
        p=0
        for k in range(0,n):
            p = p + (MAM[i][k])*(MDM[k][j])
        Mint[i][j]=p

print()
print('MATRIZ INTERMEDIÁRIA 3.28')
print()
print_matriz(Mint)

#MONTAR MATRIZ DE TRANSFERÊNCIA DE CARGA 

MC=np.zeros((n,n))

for i in range(0,n):
    for j in range(0,n):
        MC[i][j]=Mint[i][j]-Mint[j][i]

print()
print('MATRIZ TRANSFERENCIA DE CARGA: 3.28')
print()
print_matriz(MC)

#CALCULAR DO INDICE DE RANDIC:

print()
print('1X = %5.5f' %(Randic(MA,GRAU)))

#CALCULAR DO ÍNDICE DE GALVEZ:

print('Gv,4 3,28 = %5.5f' %(galvez(4, MD, MC)))
print('Jv,2 3,28 = %5.5f' %((galvez(2,MD,MC))/(n - 1))) 
print('Jv,5 3,28 = %5.5f' %((galvez(5,MD,MC))/(n - 1)))

#MONTAR MATRIZ ADJACÊNCIA MODIFICADA:

for i in range(0,len(MAM)):
    for j in range (0,len(MAM)):
        if i == j and MAM[i][j]!=0:
            MAM[i][j]= 2.2*(MAM[i][j])/3.28
        else:
            MAM[i][j]=MAM[i][j]
print()
print('MATRIZ DE ADJACÊNCIA MODIFICADA 2,2:')
print()
print_matriz(MAM)

#MONTAR MATRIZ DE ADJACÊNCIA:

MA=np.zeros((n,n))
for i in range(0,n):
    for j in range(0,n):
        if i!=j:
            MA[i][j]=MAM[i][j]
        else:
            MA[i][j]=0

#MONTAR DICIONARIO DE LISTA DE ADJACENCIA:

M=np.zeros((n,n))
for i in range(0,n):
 p=0
 for j in range(0,n):
  if MA[i][j]!=0:
      M[i][p]=j
      p=p+1
          
print()

dicio={}
lista=[]
for i in range(0,n):
    for j in range(0,n):
       if M[i][j]!=0 or j==0:
          lista.append(M[i][j])
       if j==n-1:
         dicio[i]=lista[:]
         lista=[]

#MONTAR MATRIZ DE DISTÂNCIA

MD=np.empty((n,n))

for i in range(0,n):
  for j in range(0,n):
        MD[i][j]=len(find_shortest_path(dicio,i,j)) -1

print()
print('MATRIZ DISTÂNCIA:')
print()
print_matriz(MD)

#MONTAR MATRIZ DISTÂNCIA MODIFICADA:

MDM=np.zeros((n,n))

for i in range(0,n):
  for j in range(0,n):
      if MD[i][j]!=0:
         MDM[i][j]= 1/((MD[i][j]))**2

#CALCULAR O GRAU DOS VERTICES:

GRAU=np.zeros(n)

for i in range (0,n):
    for j in range(0,n):
      GRAU[i]= GRAU[i]+MA[i][j]

#MUlTIPLICAR A MATRIZ CRIANDO MATRIZ INTERMEDIARIA: 

Mint=np.empty((n,n))

for i in range(0,n):
    p=0
    for j in range(0,n):
        p=0
        for k in range(0,n):
            p = p + (MAM[i][k])*(MDM[k][j])
        Mint[i][j]=p

print()
print('MATRIZ INTERMEDIÁRIA 2.2')
print()
print_matriz(Mint)

#MONTAR MATRIZ DE TRANSFERÊNCIA DE CARGA 

MC=np.zeros((n,n))

for i in range(0,n):
    for j in range(0,n):
        MC[i][j]=Mint[i][j]-Mint[j][i]

print()
print('MATRIZ TRANSFERENCIA DE CARGA:2,2')
print()
print_matriz(MC)

#CALCULAR O INDICE DE RANDIC:

print()
print('1X = %5.5f' %(Randic(MA,GRAU)))

#CALCULAR O ÍNDICE DE GALVEZ:

print($'Gv,4 2,2 = \%5.5f' \%(galvez(4,MD,MC))$)
print($'Jv,2 2,2 = %5.5f' %((galvez(2,MD,MC))/(n - 1))$$) 
print($'Jv,5 2,2 = \%5.5f' \%((galvez(5,MD,MC))/(n - 1))$$)}
\end{verbatim}

\end{document}